\documentclass[12pt]{article}
\usepackage{amsmath,amssymb,epsfig,graphicx}

\def\K{{\mathcal K}}

\def\be{\begin{equation}}
\def\ee{\end{equation}}

\def\K{{\mathcal K}}
\def\be{\begin{equation}}
\def\ee{\end{equation}}

\def\beq{\begin{eqnarray}}\def\eeq{\end{eqnarray}}

\begin{document}
\title{Entanglement entropy from surface terms in general relativity}
\author{Arpan Bhattacharyya  and Aninda Sinha\\
{\it email:} arpan@cts.iisc.ernet.in, asinha@cts.iisc.ernet.in \\
\it $^1$Centre for High Energy Physics,\\ \it Indian Institute of Science,  Bangalore 560 012, India.\\}

\maketitle
\vskip 2cm
\begin{abstract}
Entanglement entropy in local quantum field theories is typically ultraviolet divergent due to short distance effects in the neighbourhood of the entangling region.  In the context of gauge/gravity duality, we show that surface terms in general relativity are able to capture this entanglement entropy. In particular, we demonstrate that for 1+1 dimensional CFTs at finite temperature whose gravity dual is the BTZ black hole, the Gibbons-Hawking-York term precisely reproduces the entanglement entropy which can be computed independently in the field theory.

\end{abstract}
\vskip 1cm
{\bf Essay awarded honourable mention in the Gravity Research Foundation 2013 Awards for Essays on Gravitation.}
\newpage
The Einstein-Hilbert action for gravity needs to be supplemented by the Gibbons-Hawking-York surface term \cite{gibbons,york} to make the variational (Dirichlet boundary value) problem well defined. Namely, the total action (in Euclidean signature) is given by
\begin{align}
\begin{split}
I_{tot}&=-\frac{1}{2\ell_{P}^{d-1}}\int d^{d+1}x \sqrt{g}\big (R-2\Lambda \big)+I_{GHY}\,,\\
I_{GHY}&=-\frac{1}{\ell_{P}^{d-1}}\int d^{d}x \sqrt{h}\,\K\,.\\
\end{split}
\end{align}
where $g$ is the determinant of the bulk metric, $R$ is the scalar curvature for the bulk space time  $\Lambda$ is the cosmological constant, $I_{GHY}$ is the Gibbons-Hawking-York surface term and $\K$ is the extrinsic curvature defined on the boundary surface with the determinant of the induced metric given by $h$. Quite remarkably, even before Gibbons and Hawking or York, Einstein had written down a first order lagrangian for gravity given by $g^{a b}(\Gamma^n_{m a}\Gamma^m_{nb}-\Gamma^m_{m n} \Gamma^n_{a b})$ which differs from $I_{tot}$ by a surface term \cite{albert}. The surface terms in general relativity are crucial not only to produce a well defined variational principle but also to produce correct black hole thermodynamics. These terms are also important to compute the correct Noether charges arising from diffeomorphism invariance \cite{wald}. Furthermore, evaluating these terms on a black hole horizon, one reproduces black hole entropy \cite{bhsurf}.

In this essay, we will argue that the Gibbons-Hawking-York surface term gives the entanglement entropy in gauge/gravity duality. Entanglement entropy is a useful non-local probe of how much the degrees of freedom in a region of spacetime are entangled with the rest. The original motivation for considering entanglement entropy was the hope that such considerations would shed light on the microscopic origin of black hole entropy \cite{Bombelli:1986rw,Srednicki:1993im}. However, entanglement entropy is a useful tool in other areas of physics also, such as condensed matter systems. Recently, it has been used to shed light on time dependent physics as well where direct computational techniques are not available.  

It is known that entanglement entropy in conformal field theories in even $d$ dimensions  take the form
\be\label{def}
S_{EE}=c_d \frac{l^{d-2}}{\epsilon^{d-2}}+O(\frac{l^{d-3}}{\epsilon^{d-3}})+a_d \log \frac{l}{\epsilon}+O((\frac{l}{\epsilon})^0)\,.
\ee
 Here $l$ is a length scale parametrizing the size of the entangling region and $\epsilon$ is a short-distance cutoff. The leading $l^{d-2}$ term gives the famous area law with a non-universal proportionality constant--when $d=2$ the leading term is the $\log$ term.
In even dimensions, the coefficient of the $\log$ term is a universal quantity typically related to a function of the conformal anomalies in the theory while in odd dimensions the $\log$ term is replaced by a constant which is considered to be a measure of the degrees of freedom \cite{myersme}.   In the context of quantum field theories, a direct computation of entanglement entropy is hard and has been possible only
in very specific examples \cite{calabrese, ch}. The leading term in $S_{EE}$ is proportional to the area and gives the famous area-law. This was the main motivation for trying to relate black hole entropy with entanglement entropy of quantum fields.

The gauge/gravity correspondence or the AdS/CFT correspondence gives a way to relate a quantum (typically conformal) field theory in $d$ dimensions to a theory of gravity in anti de Sitter backgrounds in $d+1$ dimensions \cite{magoo}. The prescription in gauge/gravity duality to compute entanglement entropy in the CFT is the following. The conformal field theory is supposed to live on the boundary of the AdS space. One considers a $t=0$ slice of this boundary. Then a spatial region $\partial \mathcal{N}$ of this boundary is considered and a minimal surface $\mathcal{M}$ extending into the bulk is found which ends on $\partial \mathcal{N}$. Ryu and Takayanagi (RT) proposed \cite{ryu} that the area of this minimal surface is the entanglement entropy. To determine this minimal surface one considers an `area functional' which is then minimized. This functional when evaluated on the black hole horizon would lead to the black hole entropy.
 At the onset we should emphasise that this is a prescription, which passes several checks, with no general proof. Unlike the Wald entropy formula which is valid for a general theory of gravity, no such analogue exists for the entanglement entropy. This is an unfortunate state of affairs which needs remedy in the near future since holographic methods are becoming a popular tool to gain intuition about physics at strong coupling \cite{however}.

For computational purposes one typically uses Einstein gravity in a weakly curved anti-de Sitter (AdS) background which corresponds to a strongly coupled conformal field theory (CFT).
According to the RT prescription, in order to derive the holographic entanglement entropy for a $d$ dimensional quantum field theory,
one has to minimize the following entropy functional on a $d-1$ dimensional hypersurface (a bulk co-dimension two surface),
\be \label{ms}
S=\frac{2\pi}{\ell^{d-1}_{P}} \int d^{d-1}x \sqrt{h}\,,
\ee
where $\ell_ P$ is the Planck length and $h$ is the induced metric on the hypersurface. The resulting surface is a minimal surface with vanishing extrinsice curvature. The gravity dual theory is simply Einstein gravity with a negative cosmological constant $\Lambda=-d(d-1)/(2L^2)$ where $L$ is the AdS radius.
 This procedure can be also  used to compute entanglement entropy for the  finite temperature field theory. It corresponds to the presence of a  black hole in the bulk space time. We will consider the BTZ black hole as the result for the corresponding $1+1$ dimensional CFT is well known. Let us consider the following metric for the (non-rotating) BTZ black hole,
\be
ds^{2}= (r^2-r_{H}^{2})dt^{2}+\frac{L^{2}}{ (r^2-r_{H}^{2})}dr^{2}+r^{2} d\phi^{2}\,,\\
\ee
where, $r=r_{H}$ is the horizon. Next we put $\phi=f(r)$ into the metric and evaluate $S$. After finding the Euler-Lagrange  equation for $f(r)$ from $S$  we can determine $f(r)$ i.e.,  how the entangling surface extends into the bulk space time (see fig. 1):
\begin{align}
\begin{split} \label{f}
f(r)&=\frac{L}{r_H}\tanh^{-1} \frac{\sqrt{r^2-r_0^2}}{\sqrt{r^2-r_H^2}}\frac{r_H}{r_0}\,,
\end{split}
\end{align}
where $r_0=r_H \coth r_H l/L^2$, $l$ being the length of the entangling surface in the field theory.
Then using this solution, assuming $r_H l/L^2 \gg 1$ which corresponds to the high temperature phase of the field theory, and evaluating $S$ one gets the following well known result for the $\log$ part of the entanglement entropy:
\be
S_{EE}=\frac{2\pi}{\ell_{P}}\int_{r_0}^{\frac{1}{\epsilon}} dr  \frac{2 r L}{\sqrt{(r^2-r_0^2)(r^2-r_H^2)}} =\frac{c}{3}\log(\frac{\beta}{\pi \epsilon}\sinh (\frac{\pi l}{\beta}))+O(\epsilon)\,.
\ee
where  $c=12\pi L/\ell_P$   is the central charge of the two dimensional CFT, $\epsilon$ is the UV cut-off, $l$ is the length of the entangling surface and $\beta=1/(L T)=2\pi L /r_H$ is identified as the  periodicity in the time coordinate which is related to the inverse temperature $T$ of the field theory. An independent calculation in 1+1d produces exactly this result \cite{calabrese} which is taken to be strong evidence for the validity of the minimal area prescription. Any proposal for the entanglement entropy should agree with this.\par
\begin{figure}[h!]
\centering
\includegraphics[width=12cm, height=8cm]{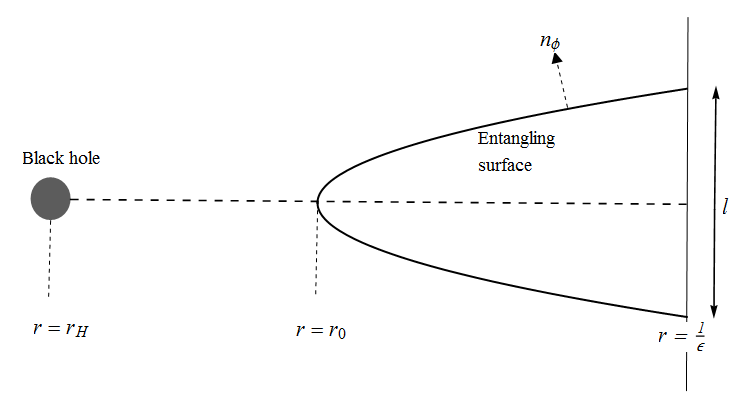}
\caption{ Entangling surface  extending into the bulk AdS. One has to add a Gibbons-Hawking-York term for both the upper and lower branch.}
 \end{figure}

Although this prescription passes certain non-trivial checks such as the strong sub-additivity condition, how does one reproduce the entangling surface in field theory? According to the AdS/CFT dictionary the AdS radius is to be identified with an RG scale. So naively one would expect that the radius of the entangling region should become a function of the RG scale. But what are the rules? Furthermore, what observable do we use in field theory to probe entanglement entropy? The RT prescription does not appear to provide direct answers to these questions. We can partially remedy this with the following observation. Consider field theory on the bulk co-dimension one slice $\phi=f(r)$. We will not set $t=0$. Since the time direction is a direct product with the rest, the trace of the extrinsic curvature satisfies $^{(d)}K_a^a= ^{(d)}\!\!\!K_t^t+ ^{(d-1)}\!\!K_i^i$. Thus $^{(d)}\!K_t^t-h_t^t \,{}^{(d)}\!K_a^a=0$ leads to $^{(d-1)}\!K_i^i=0$ which is the same as the minimal surface condition for the $d-1$ slice used in the RT calculation. But the combination $^{(d)}\!K_t^t-h_t^t \,{}^{(d)}\!K_a^a$ is nothing but the $tt$ component of the usual Brown-York (holographic \cite{stress}) stress tensor evaluated on the co-dimension one slice in $d$-dimensions! Here $h_{tt}$ is simply the $tt$ component of the pullback  metric on the co-dimension one slice--in the Brown-York tensor the indices are raised and lowered using this metric.

Thus an alternate way to compute entanglement entropy in gauge/gravity duality presents itself  \cite{Bhattacharyya:2013sia}. We first compute the time-time component of the Brown-York stress tensor on the co-dimension one entangling surface given by $\phi=f(r)$. Set this to zero and determine $f(r)$. The above argument guarantees that $f(r)$ will work out to be the same as what follows from the RT prescription. Now it is intuitive, that since entanglement entropy is related to the common boundary between the degrees of freedom living in the two regions, one of which is traced over, it must be related to the surface terms in general relativity. Now recall that the RT area functional was such that evaluated on the black hole horizon, we got the black hole entropy. The Gibbons-Hawking-York term evaluated on the horizon of a black hole is also known to yield  black hole entropy \cite{bhsurf}. Let us work out the Gibbons-Hawking-York surface term in our case explicitly. Unlike the RT area functional, there is a time integral in this case. But we know that time has to be periodic, with period $\beta=1/T$, $T$ being the temperature of the BTZ black hole. After some straightforward algebra, we get

\begin{align}
\begin{split}
I_{GHY}&=\frac{1}{\ell_{P}}\int_{0}^{\beta=\frac{2\pi L }{r_{H}}} dt \int_{r_0}^{\frac{1}{\epsilon}} dr \frac{2 r r_0}{\sqrt{(r^2-r_0^2)(r^2-r_H^2)}}\,,\\
&=\frac{c}{3}\log(\frac{\beta}{\pi \epsilon}\sinh (\frac{\pi l}{\beta}))+O(\epsilon)\,,\end{split}
\end{align}
where in going to the second line we have assumed that the field theory is in the high temperature phase which makes $r_0\approx r_H$. Thus the RT result is identical to what comes from the Gibbons-Hawking-York surface term. This agreement can be shown to hold even at zero temperatures if one makes time periodic with the periodicity related to the inverse Unruh temperature. Also this connection holds for any dimensions not just 1+1d. For the computations in higher dimensions readers are referred to \cite{Bhattacharyya:2013sia,Bhattacharyya:2013jma}\,. This method can also be applied for the stationary cases such as  rotating BTZ. The explicitly time-dependent situations are left for future work.
\vskip 0.5 cm
\noindent {\bf Conclusions}\\
We have shown that entanglement entropy for field theories having holographic duals are related to the surface terms arising in general relativity. This may point at a more systematic way of computing entanglement entropy by relating it to Noether charges in general relativity. Since the procedure for computing entanglement entropy was given in terms of the Brown-York stress tensor, this naturally suggests a possible way to find out about the entangling surface using field theory methods. Some evidence for this has been presented in \cite{Bhattacharyya:2013sia}.

\vskip 0.5 cm
\noindent{ \bf Acknowledgments}\\
We thank Gautam Mandal, Rob Myers and Tadashi Takayanagi for useful discussions and correspondence.

\end{document}